\documentclass{aa}
\usepackage[varg]{txfonts}

\def\apjl{ApJ}
\def\aap{A\&A}

\def\Msun{M$_\odot$}
\def\Rsun{R$_\odot$}
\def\ms{m~s$^{-1}$}
\begin{document}
\title{Parallax and masses of $\alpha$ Centauri revisited\thanks{Based on data obtained from the ESO Science Archive Faci\-lity under request numbers HBOFFIN/190700, Pourbaix/192287, Pourbaix/192364, Pourbaix/192404, Pourbaix/192552, Pourbaix/192630, and Pourbaix/199124.}}

\author{Dimitri~Pourbaix\inst{1}\fnmsep\thanks{Senior Research Associate, F.R.S.-FNRS
, Belgium}
\and Henri~M.~J.~Boffin\inst{2}}
\institute{Institut d'Astronomie et d'Astrophysique, Universit\'e Libre de Bruxelles (ULB), Belgium\\
\email{pourbaix@astro.ulb.ac.be}
\and ESO, Alonso de C\'ordova 3107 Vitacura, Casilla 19001 Santiago, Chile}
\date{Received 30/11/2015; accepted 04/01/2016}
\AANum{AA/2015/27859}
\abstract{Despite the thorough work of van Leeuwen (2007), the parallax of $\alpha$ Centauri is still far from being carved in stone.  Any derivation of the individual masses is therefore uncertain, if not questionable.  And yet, that does not prevent this system from being used for calibration purpose in several studies.}
{Obtaining more accurate model-free parallax and individual masses of this system.}
{With HARPS, the radial velocities are not only precise but also accurate.  Ten years of HARPS data are enough to derive the complement of the visual orbit for a full 3D orbit of $\alpha$ Cen.}
{We locate $\alpha$ Cen (743 mas) right where Hipparcos (ESA 1997) had put it, i.e. slightly further away than derived by S\"oderhjelm (1999).  The components are thus a bit more massive than previously thought (1.13 and 0.97 \Msun\ for A and B respectively).  These values are now in excellent agreement with the latest asteroseismologic results.}
{}


\maketitle

\section{Introduction}\label{sec:intro}
The Sun is a single star and as such is among the minority of solar-like stars which are mostly within binaries or multiple systems \citep{Duquennoy-1991:b,Halbwachs-2003:a,Raghavan-2010:a,Whitworth-2015:a}. 
Our closest neighbour -- the system comprising $\alpha$ Centauri A, B and Proxima Centauri -- is therefore more representative.
\object{$\alpha$ Centauri} A and B (HIP~71683/1), with spectral types G2\,V and K1\,V,  are in a binary system with an orbital period close to 79.91 years \citep{Heintz-1982:b,Pourbaix-1999:b,Torres-2010}  and a distance of 1.35 pc. The A and B pair offers a unique possibility to study stellar physics in stars that are only slightly different from our own Sun. Their masses -- 1.1 and 0.9 Msun -- nicely bracket that of our neighbour star, and they are only slightly older than the Sun. Thus, $\alpha$ Cen is an ideal laboratory for stellar evolution \citep[e.g.][]{Kervella-2003:a,PortodeMello-2008:a,Bruntt-2010:a,Bazot-2012:a}, asteroseismology \citep{Kjeldsen-2008:a,deMeulenaer-2010:a} and extra-solar planet searches \citep{Dumusque-2012:a,Rajpaul-2015:a,Bergmann-2015:a,Endl-2015:a}. As such it is crucial to determine with the highest accuracy the properties of the two stars in $\alpha$ Cen, which can be done as double-lined spectroscopic  visual  binaries, thus offering a hypothesis-free determination of the distance and individual masses \citep{Pourbaix-1998:a}. One also needs to have the most precise orbital elements to disentangle any other effects, such as oscillations or the presence of a planetary-mass companion.

\citet{Pourbaix-1999:b} presented the first simultaneous adjustment of the relative positions and radial velocities of both components of $\alpha$ Cen, yielding an upward revision of the masses.  Owing to the special interest of the asteroseismology  community for this system, an international team was gathered later on to obtain some accurate radial velocities of both components.  The outcome was a set precise radial velocities which were used to quantify the relative convective blue shift of both components, assuming the individual masses and the parallax of the system \citep{Pourbaix-2002:a}. 

Even if it turned out to be a false detection \citep{Hatzes-2013:a,Rajpaul-2015:a}, the announcement of a planetary companion around $\alpha$ Cen B \citep{Dumusque-2012:a} suddenly resurrected the interest of the planet hunters for that stellar system \citep{Kaltenegger-2013:a}. We therefore decided to determine for the first time, in a self-consistent manner, the individual masses, the parallax, and the net shift caused by gravitation and convection, using an extensive set of homogeneous and accurate radial velocities of both components from the ESO HARPS science archive. The observations are described in Sect.~\ref{sec:observations} while the adopted model used to fit them is described in Sect.~\ref{sec:model}.  Results are listed in Sect.~\ref{sec:results} and discussed in the context of asteroseismology in Sect.~\ref{sec:discussion}.

\section{Observational data}\label{sec:observations}
$\alpha$~Cen has been the target of many radial velocities (RV) measurements, especially, with HARPS, the High Accuracy Radial velocity Planet Searcher at the ESO La Silla 3.6m telescope. The vacuum and thermally isolated HARPS instrument has been especially designed for high-precision radial velocities observations \citep{Mayor-2003:a}, reaching for example a dispersion of 0.64 m~s$^{-1}$ over 500 days \citep{Lovis-2006:a}.

The velocities of both components of  $\alpha$~Cen were retrieved from the HARPS archive maintained by ESO: 2015 velocities for A and 4303 for B.  Despite the possibility of selecting the target on the ESO archive interface, a visual inspection was necessary to assign the velocities to the right component.  Further imposing that the seeing does not exceed 1 arcsec so as to avoid $\alpha$ Cen A contaminating $\alpha$ Cen B and vice versa (as suggested by the referee, Xavier Dumusque), limited these observations to 710 and 1951 for A and B respectively.  The importance of this data set lies in the simultaneous or quasi simultaneous observations of both components with an instrument that provides RVs on an almost absolute scale. 

\begin{figure}[ht]
\begin{center}
\resizebox{\hsize}{!}{\includegraphics{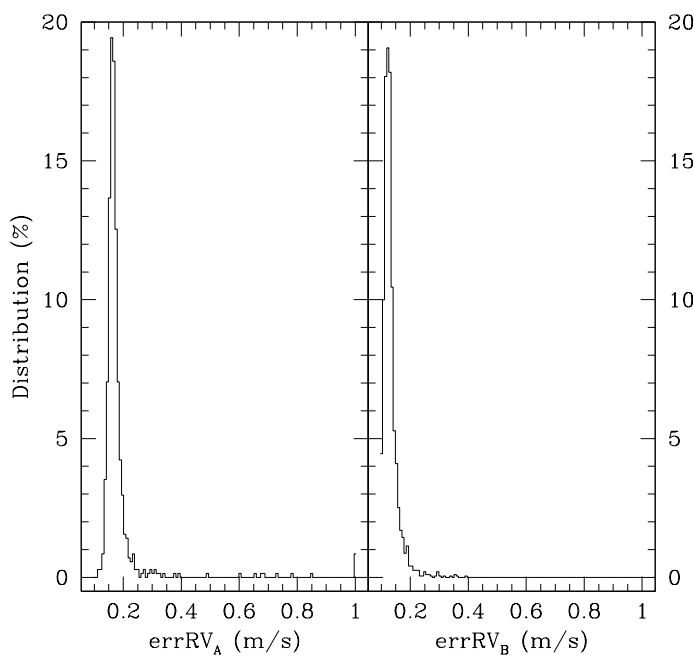}}
\end{center}
\caption[]{\label{fig:ccferr}Distribution of the estimated radial velocity uncertainties reported by the HARPS pipeline for component A (left) and B (right).}
\end{figure}

We used the radial velocities provided by the HARPS pipe\-line.  For $\alpha$~Cen A, the RV is obtained by cross-correlating the spectra with a G2 V flux template which is the Fourier transform spectrometer (FTS) spectrum of the Sun \citep{SoFlAt}, and calibrated so as to have an offset in the zero-point of 102.5 \ms\ \citep{Molaro-2013:a}. For $\alpha$~Cen B, the cross-correlation was done with a K5 template.  The median of the velocity precision for A and B are respectively 0.16 and 0.12 \ms\ (Fig.~\ref{fig:ccferr}).

The HARPS data cover 11 years only (13\% of the orbital period), but at the crucial time when the radial velocities cross (see Fig.~\ref{fig:plot}). The HARPS data were completed with some older ESO data \citep{Endl-2001:a}, obtained with the Coud\'e Echelle Spectrograph (CES) at the 1.4-m Coud\'e Auxiliary Telescope and later, at the 3.6-m telescope, both in La Silla, to extend the baseline and to help improving the precision of the fractional mass ($\kappa=M_B/(M_A+M_B)$).  These velocities being relative, the datasets of A and B were shifted to share the HARPS zero point.

These very accurate radial velocities were complemented with the same visual observations (both micrometric and photographic) as used in our previous investigation \citep{Pourbaix-1999:b}.  According to its web portal, the Washington double star catalogue \citep{Hartkopf-2001:b} holds 37 additional visual observations (up to 2014.241) with respect to our original investigation.  These data were kindly provided by the WDS team and added to the 1999 dataset for the sake of completeness.  In practice, no parameter from the visual orbit was affected.

\section{Model}\label{sec:model}

The model used by \citet{Pourbaix-1999:b} assumes that the measured radial velocities represent the radial velocities of the barycentre of each component:
\begin{equation}\label{eq:rvB}
  \begin{array}{l}
  V_A=V_0-K_A(e\cos\omega_B+\cos(\omega_B+v)),\\
  V_B=V_0+K_B(e\cos\omega_B+\cos(\omega_B+v)),
  \end{array}
\end{equation}
where $V_0$ denotes the systemic velocity, $\omega_B$ the argument of the periastron of component B, $e$ the eccentricity,  $v$ the true anomaly, and $K_{A,B}$ are the semi-amplitudes of the radial velocities of both components.

Whereas that assumption was realistic in the past when the radial velocities were precise to a few hundred meters per second, some effects pop up as soon as the precision improves.  In order to recover the accuracy of the barycentre velocity, these effects have to be corrected for, either individually or globally.  With relative radial velocities of both components, these effects would have to be modelled.  With HARPS measuring both components in the same reference frame, it is possible to measure the correction to be applied globally. 

Assuming the gravitational red shift and convective blue shift of a given component do not change over the spectroscopic observation baseline, the net effect of the two shifts is just a vertical translation of the radial velocity curve (the dates of the minimums and maximums of the curve remain unchanged).  No morphological change of the curve itself is anticipated.  The net effect of the four shifts is therefore a vertical translation of one curve with respect to the other.

Such a vertical translation can easily be modelled with an additional term in, say, the radial velocity of component B (Eq.~\ref{eq:rvB}):
\begin{equation}
  V_B=V_0+K_B(e\cos\omega_B+\cos(\omega_B+v))+\Delta V_B.\label{eq:rvBnew}  
\end{equation}
It is worth pointing that, whereas \citet{Pourbaix-1998:a} advocated for a simultaneous adjustment of the visual and spectroscopic data, this $\Delta V_B$ term has to be introduced because the solutions for $V_A$ and $V_B$ are obtained simultaneously!  Indeed, if the two curves were modelled independently, two distinct $V_0$ would be obtained but $K_A$ and $K_B$ would represent the semi-amplitudes of the two curves.  Without $\Delta V_B$, the simultaneous fit introduces a bias on $V_0$, $K_A$, and  $K_B$.

The orbit of the stellar system being our only goal, no short timescale variation \citep{Dumusque-2011:a,Dumusque-2015:a} is modelled in the present investigation.

\section{Results}\label{sec:results}
\begin{figure}[ht]
\begin{center}
\resizebox{\hsize}{!}{\includegraphics{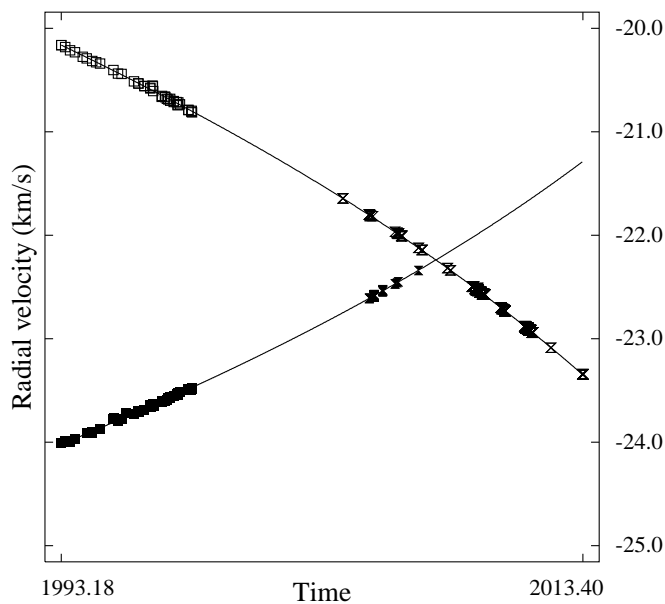}}
\end{center}
\caption[]{\label{fig:plot}Radial velocities of alpha Cen (filled for component A and open for B).  Diabolos denote the HARPS archived data and squares the older ESO data \citep{Endl-2001:a} already used by \citet{Pourbaix-2002:a}.  On this portion of the orbit, fitting $\Delta\ V_B$ or setting it to 0 is not visually distinguishable.}
\end{figure}

Despite the absence of visual departure between the fit of the present dataset with and without $\Delta\ V_B$, the parallaxes differ by 2\% (smaller without shift), directly impacting the total mass by the same amount as the fractional mass remains essentially unchanged.  The reduced $\chi^2$ increases from 1.01 to 1.21 without the shift. The revision of the model is thus justified.  The orbital elements are given in Tab.~\ref{tab:orbit} together with the 2002 results and the orbit is plotted in Fig.~\ref{fig:plot}.

The revised orbital parallax ($743\pm1.3$ mas) is smaller than the value derived by \citet{Soderhjelm-1999:b} from the Hipparcos observations and adopted by \citet{Pourbaix-2002:a}.  It is somewhat closer to the original Hipparcos value, $742\pm1.42$ mas \citep{Hipparcos}, and rules out the result obtained in the revision of the Hipparcos catalogue \citep{Hip2} where the parallax is $754.81\pm4.11$ mas.  Even though the parallax is different, the total mass of the system perfectly matches the 'photometric' estimate from \citet{Malkov-2012:a}, thus indicating some possible flaw in their mass-luminosity relation.  Our value of the mass of component B seems to favour the asteroseismology-based $0.97\pm0.04$ \Msun\ by \citet{Lundkvist-2014:a} over the 0.921 \Msun\ based on isochrone interpolation \citep{Boyajian-2013:a}.

\begin{table}[htb]
  \caption[]{\label{tab:orbit}Orbital solutions from \citet{Pourbaix-2002:a}, this work using HARPS and some older ESO Coud\'e Echelle velocities \citep{Endl-2001:a}.}
  \begin{tabular}{lcc}
    \hline\hline
            &          & HARPS + ESO\\
    Element & Original & Coud\'e Echelle \\
    \hline
$a$ (\arcsec)       & $17.57\pm0.022$          &$17.66\pm0.026$\\
$i$ (\degr)         & $79.20\pm0.041$          &$79.32\pm0.044$\\
$\omega$ (\degr)    & $231.65\pm0.076$         &$232.3\pm0.11$\\
$\Omega$ (\degr)    & $204.85\pm0.084$         &$204.75\pm0.087$\\
$e$                 & $0.5179\pm0.00076$       &$0.524\pm0.0011$\\
$P$ (yr)            & $79.91\pm0.011$          &$79.91\pm0.013$\\
$T$ (Julian year)   & $1875.66\pm0.012$        &$1955.66\pm0.014$\\
$V_0$ (km\,s)       & $-22.445\pm0.0021$       &$-22.390\pm0.0042$\\
$\varpi$ (mas)      & $747.1\pm1.2$  (adopted) &$743\pm1.3$\\
$\kappa$            & $0.4581\pm0.00098$       &$0.4617\pm0.00044$\\
\medskip
$\Delta V_B$ (m/s ) & 0.0 (adopted)             &$329\pm9.0$\\
$M_A$ (\Msun)       & $1.105\pm0.0070$          &$1.133\pm0.0050$\\
$M_B$ (\Msun)       & $0.934\pm0.0061$          &$0.972\pm0.0045$\\
    \hline
    \end{tabular}
\end{table}

In the particular case of this system, $\Delta V_B$ can be interpreted as the net effect, for component B only, of the differential gravitational redshift, differential convective blue shift and template mismatch.  Indeed, the template used for component A is a G2 mask calibrated against asteroids.  The radial velocities of A are therefore as close to absolute as possible.  Component B was reduced using a K5 mask instead of K1 (the commonly accepted spectral type).

\begin{figure}[ht]
\begin{center}
\resizebox{\hsize}{!}{\includegraphics{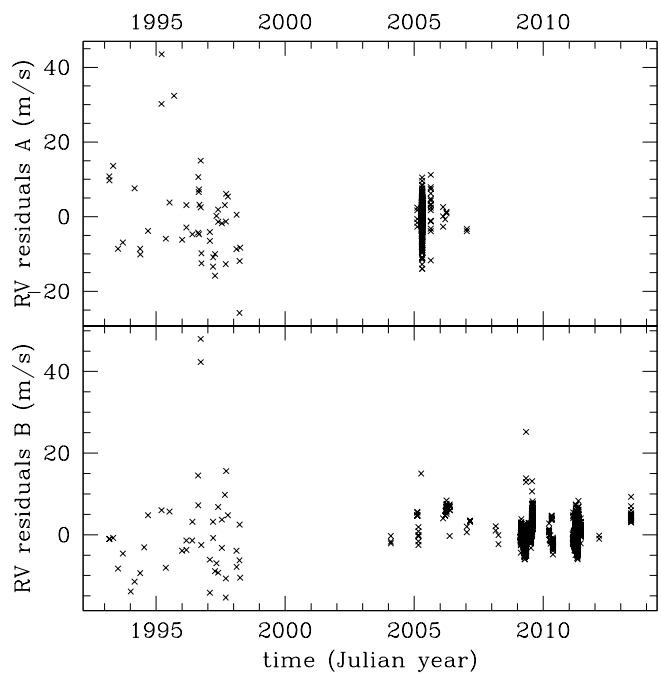}}
\end{center}
\caption[]{\label{fig:residuals}Radial velocity residuals of both components (top: A; bottom: B) resulting from the orbital fit.  The HARPS data are all located after 2000.  The older data are from ESO Coud\'e Echelle \citep{Endl-2001:a}.}
\end{figure}

The velocity residuals against the orbit have a standard deviation of 4.56 and 3.26 \ms\ for A and B respectively (Fig.~\ref{fig:residuals}).  For the HARPS data only, the standard deviations are 3.44 and 2.74 \ms\ with the latter likely overestimated due to some outliers in 2009 not filtered out by the constraint on the seeing.  Those values, especially for B, are consistent with the residuals obtained by \citet{Dumusque-2012:a} before they corrected for other effects (e.g. rotational activity, \dots).

\section{Discussion}\label{sec:discussion}
\citet{Thevenin-2002:a} could not find any asteroseismologic model consistent with the masses obtained by \citet{Pourbaix-2002:a} and, instead, proposed $1.100\pm0.006$ \Msun\ and $0.907\pm0.006$ \Msun\ for $\alpha$ Cen A and B respectively.  These results were somehow confirmed by \citet{Kervella-2003:a} through the measurement of the angular diameter of both components and adopting the parallax by \citet{Soderhjelm-1999:b}.  Combining their own results with those of \citet{Thevenin-2002:a}, they also derived a likely parallax of $745.3\pm2.5$ mas.

Using asteroseismology only, \citet{Lundkvist-2014:a} obtained $1.10\pm0.03$ \Msun\ and $0.97\pm0.04$ \Msun, very consistent with our values.  They also derived $1.22\pm0.01$ \Rsun\ and $0.88\pm0.01$ \Rsun\ for the radius of component A and B respectively, matching the values obtained by \citet{Kervella-2003:a}.  Adopting the angular diameters from the latter ($8.511\pm0.020$ mas and $6.001\pm0.034$ mas for A and B) and our revised parallax yield $1.231\pm0.0036$ \Rsun\ and $0.868\pm0.0052$ \Rsun\ for the radii of A and B, also in good agreement with \citet{Lundkvist-2014:a}.

\section{Conclusions}\label{sec:conclusions}
As stressed by several authors \citep{Torres-2010,Halbwachs-2016:a}, obtaining stellar masses at the 1\% level is crucial for astrophysics.  Accounting for $\Delta V_B$ made it possible to reach that level of precision (and hopefully of accuracy as well) for $\alpha$ Cen without any ad-hoc assumption over a so short timescale.  The revised distance and masses match the values independently derived by astroseismology.

\begin{acknowledgements}
We thank the referee, Xavier Dumusque, for his suggestion about filtering the HARPS dataset according to the seeing.  This research has made use of the Washington Double Star Catalog maintained at the U.S. Naval Observatory and the Simbad data base, operating at CDS, Strasbourg, France.
\end{acknowledgements}

\bibliographystyle{aa}
\bibliography{articles,books}

\end{document}